\documentclass{WileyMSP-template}

\usepackage{color,soul}

\usepackage{svg}

\usepackage{pdfpages}

\begin{document}

\pagestyle{fancy}
\lhead{}
\rhead{}

\title{Influence of Dimensionality of Carbon-based Additives on Thermoelectric Transport Parameters in Polymer Electrolytes}

\maketitle

\author{Maximilian Frank*}
\author{Julian-Steven Schilling}
\author{Theresa Zorn}
\author{Philipp Kessler}
\author{Stephanie Bachmann}
\author{Ann-Christin Pöppler*}
\author{Jens Pflaum*}

\dedication{}

\begin{affiliations}
Maximilian Frank, Julian-Steven Schilling, Prof. Dr. Jens Pflaum\\
Experimental Physics VI, Julius-Maximilian University, D-97074 Würzburg, Germany\\
Email Address: maximilian.frank@uni-wuerzburg.de, jens.pflaum@uni-wuerzburg.de

Theresa Zorn, Philipp Kessler, Stephanie Bachmann, Prof. Dr. Ann-Christin Pöppler\\
Institute of Organic Chemistry, Center for Nanosystems Chemistry Julius-Maximilian University, D-97074 Würzburg, Germany\\
Email Address: ann-christin.poeppler@uni-wuerzburg.de

Prof. Dr. Jens Pflaum\\
Center for Applied Energy Research e.V. (CAE), D-97074 Würzburg, Germany

\end{affiliations}


\begin{abstract}
\justifying
This paper investigates the thermoelectric properties of solid polymer electrolytes (SPE) containing lithium bis(tri\-fluoro\-methane)\-sulfonimide (LiTFSI) and sodium bis(trifluoromethanesulfonyl)imide (NaTFSI) salts, respectively, as well as carbon-based additives of various dimensionalities. By means of experimental characterization and analysis, our study reveals that increasing salt concentration leads to higher Seebeck coefficients, as a result of the increasing number of free charge carriers and additional, superimposed effects by ion-ion and ion-polymer interactions. NaTFSI-based electrolytes exhibit negative Seebeck coefficients of up to $S = -1.5\,\mathrm{mV\,K^{-1}}$, indicating dominant mobility of $\mathrm{TFSI^-}$ ions. Carbon nanotubes (CNTs) of quasi one-dimensional character significantly increase the Seebeck coefficient by a factor of $3$. Planar, 2D graphite flakes (GF) moderately enhance the Seebeck coefficient by affecting $\mathrm{Na^+}$ and $\mathrm{TFSI^-}$ ion mobilities, and furthermore contribute to electronic conductivity. Bulky, 3D carbon Black (CB) additives induce a unique behavior where the sign of the Seebeck coefficient changes with temperature, presumably due to interaction with $\mathrm{TFSI^-}$ ions within the CB structure. Complementary, our study also hints at changes in activation energy and Vogel temperature with salt concentration, indicating structural and mechanical modifications in the polymer matrix. Overall, the choice of carbon-based additives and salt concentration significantly influences the thermoelectric properties of SPEs, providing insights into their potential for thermoelectric applications. Furthermore, sodium-based electrolytes show great promise as sustainable alternatives to lithium-based systems, highlighting their technological significance for thermoelectric and energy storage applications and aligning with current demands in sustainable energy research.

\end{abstract}

\justifying

\section{Introduction}
Polymer electrolytes have emerged as a promising class of materials for energy storage and conversion applications, offering enhanced mechanical flexibility, improved safety, a wide range of electrochemical stability and processability  $^{[1]}$. Their ionic transport characteristics enable intriguing utilization in thermoelectrics (TE) to recover substantial amounts of waste heat into electrical power $^{[2]}$.
Sustainable energy conversion solutions are increasingly sought after, with a particular focus on direct conversion of waste heat into usable electrical power. The high costs and toxicity associated with conventional solid-state thermoelectric materials have hindered large-scale application, so far. Hence, overcoming these limitations offers important technological prospects.
Efficient TE applications require materials with large Seebeck coefficient, $S$, high electrical conductivity, $\sigma$, and low thermal conductivity, $\kappa$. These parameters are commonly interrelated by the dimensionless figure-of-merit at a given temperature $zT = S^2 \sigma T \kappa^{-1}$. To achieve high power factors, defined as $PF = S^2 \sigma$ $^{[3]}$, in purely electronic systems, however, the two most important parameters of TE transport, $S$ and $\sigma$, are quite often in conflict as they are anticorrelated. Despite tremendous efforts and various approaches, like doping, incorporation of additives such as low dimensional molecular metals, organic charge transfer salts or carbon nanotubes, the Seebeck coefficient in polymer materials with neat electronic conductivity remains quite low, only a few $10\,\mathrm{\mu V\,K^{-1}}$ $^{[4][5][6][7][8][9]}$. The main reason for this limitation is given by the dependence of $S$ on the charge carrier concentration. In materials with ionic transport, however, the lower charge carrier mobility and the higher effective mass causes lower conductivity and therefore enhanced Seebeck coefficients $S$. As such, solid polymer electrolytes (SPE) with thermovoltages in the mV range would offer an excellent alternative to materials showing only electronic transport $^{[10][11]}$. At the same time as thermoelectric properties are being improved, the sustainability of electrolytes and alternatives to lithium are becoming increasingly important, thus cross-fertilizing the field of polymer electrolytes for sustainable energy applications.
In extension to our previous studies on lithium-based solid-state electrolytes, where we investigated the concentration dependent thermoelectric properties $^{[12]}$, here, we present a comparative study on sodium-based solution processable polymer electrolytes, further investigating the underlying mechanisms governing charge carrier transport and extending the understanding of the interplay between structural and thermoelectric properties.
By adding carbon-based materials commonly used in battery research, we investigate the influence of the additive’s dimensionality on the key transport properties. 
This approach leads to high Seebeck coefficients of about $2\,\mathrm{\mu V\,K^{-1}}$ on macroscopic scales of several millimeters, which refer to realistic dimensions of thermoelectric generators (TEG) and are necessary to maintain a decent temperature gradient $^{[13]}$. Through our findings, we shed light on the critical factors influencing charge transport and ion conductivity within the polymer electrolyte matrix and showcase the huge potential inherent to polymer electrolytes for high performance thermoelectric applications.

\section{Results and Discussion}
In accordance with the investigations by our group on lithium-containing electrolytes $^{[12]}$, we have employed the solid polymer electrolyte poly(ethylene glycol)-methacrylate (PEGMA) : bisphenol-A-ethoxylate-dimethacrylate (BEMA) in a weight-to-weight ratio of 3:7 as the basic material for our studies presented here. This choice has been made with the specific intention of carrying out a comparative analysis of present and previous findings. Notably, this polymer electrolyte exhibits a remarkable thermal stability up to $300\,\mathrm{^{\circ}C}$ $^{[14]}$, rendering it suitable for thermoelectric generators designed to recover waste heat within the technologically significant mid-temperature range $^{[2]}$. Moreover, from a preparational point of view, the viscosity of the unhardened electrolyte offers printing-based fabrication and, in combination with the swift UV-curing of the polymer, access to easy up-scaling.
In our previous study we investigated the influence of variable concentrations of lithium bis(trifluoromethane)-sulfonimide (LiTFSI) on the thermoelectric properties. Given the growing significance of sustainable energy production, there is a strong imperative to transition from lithium to sodium $^{[15][16]}$. Sodium possesses a considerably more abundant presence across the Earth's crust and offers the advantage of a more environmentally friendly extraction process. Consequently, these circumstances have also stimulated our research activities reported here where we investigate the thermoelectric performance of polymer electrolytes as function of sodium bis(trifluoromethane)-sulfonimide (NaTFSI) concentration and carbon-based additivies of different dimensions.

\subsection{Thermoelectric and Electrical Characterization of the Host Electrolyte}
\begin{figure}[!h]
  \includegraphics[width=\linewidth]{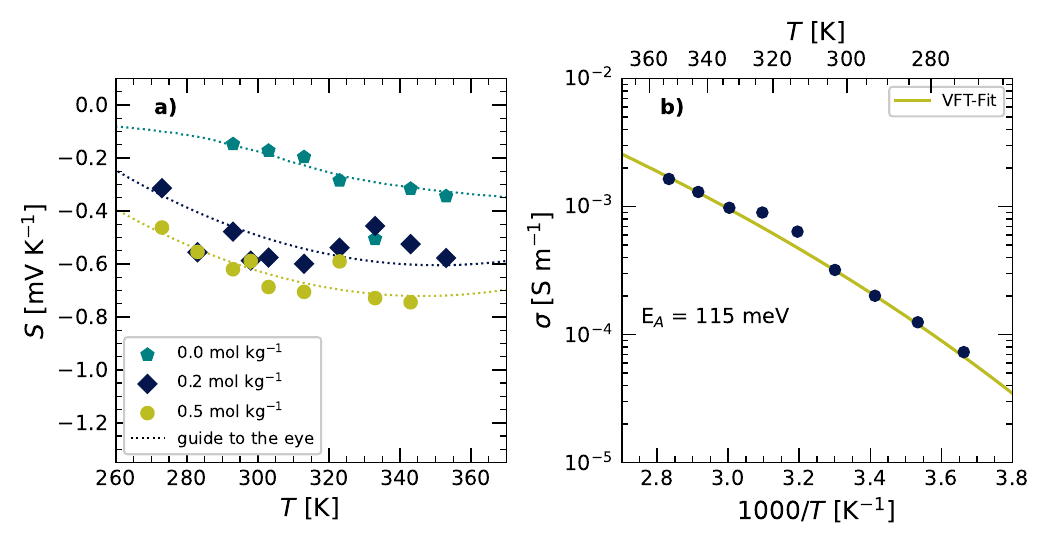}
  \caption{\textbf{a)} Seebeck coefficient $S$ as function of device temperature for different salt concentrations $c_{\mathrm{NaTFSI}}\,=\,0.0\,\mathrm{mol\,kg^{-1}}$ (Na0, teal pentagons), $c_{\mathrm{NaTFSI}}\,=\,0.2\,\mathrm{mol\,kg^{-1}}$ (Na2, blue diamonds) and $c_{\mathrm{NaTFSI}}\,=\,0.5\,\mathrm{mol\,kg^{-1}}$ (Na5, green dots),  respectively. Data for sample Na0 were taken from $^{[12]}$. \textbf{b)} Ionic conductivities in the polymer matrix derived from impedance spectra (blue dots). The temperature dependent conductivity can be described by a Vogel-Fulcher-Tammann (VFT) ansatz (green solid line), yielding an activation energy of $E_A\,=\,\left(115\,\pm\, 11 \right)\,\mathrm{meV}$ and a Vogel temperature of $T_{V}\,=\,\left( 127\,\pm\,21\right)\,\mathrm{K}$.}
  \label{fig:seebeck}
\end{figure}

\begin{figure}[!h]
  \includegraphics[width=\linewidth]{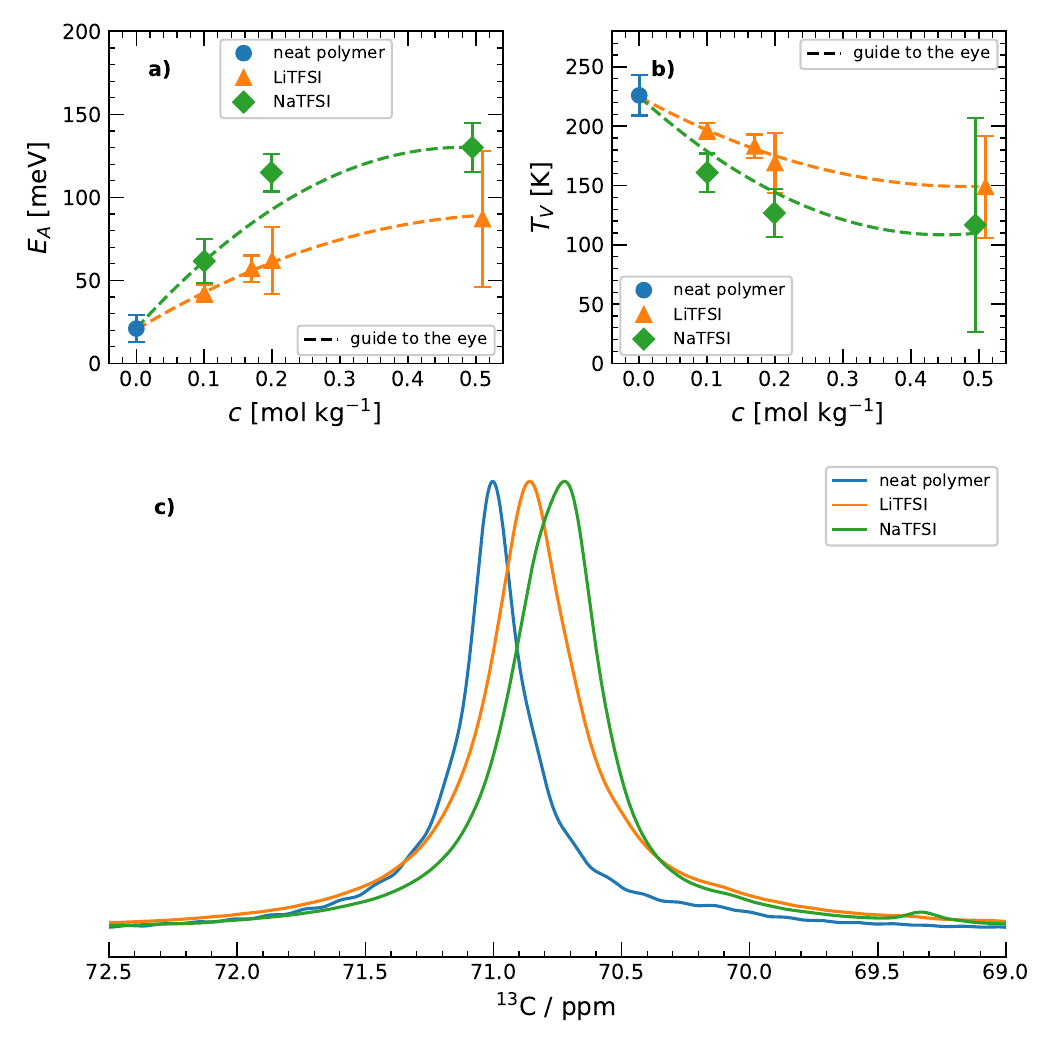}
  \caption{\textbf{a)} Activation energy $E_A$ and \textbf{b)} Vogel temperature $T_V$ according to VFT formalism as a function of salt concentration. The blue dots indicate the respective value for the neat polymer. While the Vogel temperature decreases with increasing salt concentration $c_{\mathrm{LiTFSI}}$ (orange triangles) and $c_{\mathrm{NaTFSI}}$ (green diamonds), respectively, the activation energy increases. \textbf{c)} Signals of the PEG-group in $^{13}$C solid-state NMR spectra obtained by direct excitation (DE) reveal the reduction of the chemical shift of the mobile component of the PEG-group at $71\,\mathrm{ppm}$ to be characteristic for the respective conductive salt. $^{13}$C hpdec with short interscan delay of $1\,\mathrm{s}$.}
  \label{fig:nmr}
\end{figure}

The thermoelectric voltage ($U_{th}$) was measured as a function of the applied temperature gradient ($\Delta T$) and salt concentration $c_{\mathrm{NaTFSI}}$ to determine the Seebeck coefficient. Three samples have been prepared with concentrations $c_{\mathrm{NaTFSI}}\,=\,0.0\,\mathrm{mol\,kg^{-1}}$ (Na0), $0.2\,\mathrm{mol\,kg^{-1}}$ (Na2) and $0.5\,\mathrm{mol\,kg^{-1}}$ (Na5). The Seebeck coefficient was obtained from the slope of the linear fit to the data. \textbf{Figure \ref{fig:seebeck}a} presents the Seebeck coefficients for the respective salt concentration ($c_{\mathrm{NaTFSI}}\,=\,0.0\,\mathrm{mol\,kg^{-1}}$ (teal pentagons), $0.2\,\mathrm{mol\,kg^{-1}}$ (blue diamonds) and $0.5\,\mathrm{mol\,kg^{-1}}$ (green dots)) as function of device temperature. Data for the neat polymer were taken from Ref. $^{[12]}$. Analyzing the data presented in Figure \ref{fig:seebeck}, it becomes evident that all examined samples manifest negative Seebeck coefficients. This observation indicates the predominant contribution, due to their higher mobility, of the negatively charged $\mathrm{TFSI^{-}}$ ions to the arising thermoelectric voltage. Consequently, it follows that the positively charged $\mathrm{Na^+}$ ions experience limitations in their mobility, likely due to coordination with carbonyl oxygen atoms of the methacrylate groups or ether oxygen atoms within the PEG blocks. \newline
In its pristine form, i.e. without any additives in the polymer matrix, sample Na0 shows a Seebeck coefficient of $S\,=\,\left(-0.14\,\pm\,0.12\right)\,\mathrm{mV\,K^{-1}}$ at room temperature $^{[12]}$, corresponding to the anticipated range for untreated poorly conducting polymers. Upon adding NaTFSI salt at a concentration of $c_{\mathrm{NaTFSI}}\,=\,0.2\,\mathrm{mol\,kg^{-1}}$ to the polymer host (sample Na2), the Seebeck coefficient increases to 
$S\,=\,-0.48\,\mathrm{mV\,K^{-1}}$. Higher salt concentration ($c_{\mathrm{NaTFSI}}\,=\,0.5\,\mathrm{mol\,kg^{-1}}$, sample Na5) results in an even more pronounced increase in the Seebeck coefficient, reaching $S\,=\,-0.62\,\mathrm{mV\,K^{-1}}$ at room temperature. As higher salt concentrations lead to an increasingly brittle, thus hardly handable polymer upon UV curing, no higher concentrations were tested. \newline
Up to a temperature of $T\,=\,323\,\mathrm{K}$, an increase in salt concentration correlates with an increase in $|S|$. Beyond this threshold, the inclination flattens, leading to a plateau in the thermovoltage. This phenomenon can be ascribed to the concentration-dependent charge carrier mobility. Higher charge carrier concentrations yield an enhanced probability for repulsive ion-ion interactions, resulting in an effective reduction in charge carrier mobility within the polymer matrix. This effect is corroborated, e.g. by the reduced ion diffusion coefficients observed for $\mathrm{LiPF_6}$ and $\mathrm{LiBF_{4}}$ in propylene carbonate solutions $^{[17]}$. \newline
At a temperature of $T\,=\,353\,\mathrm{K}$, the absolute value of the Seebeck coefficient for sample Na5 ($S\,=\,-0.44\,\mathrm{mV\,K^{-1}}$, green dots) is smaller than that of Na2 ($S\,=\,-0.53\,\mathrm{mV\,K^{-1}}$) which can be attributed to increased ion-ion interaction at higher concentrations and temperatures, restricting their mobility. Overall, we constitute a non-linear relationship between the Seebeck coefficient and salt concentration in the polymer electrolyte. Higher salt concentration boosts the Seebeck coefficient due to the increasing number of free charge carriers, facilitating charge transport in the polymer matrix (\textbf{Supp. Figure 1} in the Supplementary Information). \newline
In comparison, LiTFSI yields higher Seebeck coefficients for the same polymer matrix and at equivalent concentrations and temperatures $^{[12]}$. As the $\mathrm{Li^+}$ ion constitutes a hard Lewis acid according to the HSAB concept of hard and soft acids and bases, it shows a higher affinity to coordinate to the carbonyl and ether sites, reducing the overall cation mobility within the polymer matrix and, consequently, making the TFSI anion diffusion the dominating process in the LiTFSI system. The differences in the ionsic mobilities lead to inbalanced concentration gradients of the different ionic species upon the application of a temperature gradient and, thus, gives rise to the thermovoltage. It should be noted that the greater the difference in mobilities, the greater the magnitude of the voltage. Thus, the resulting Seebeck coefficient is increased for the LiTFSI system compared to the NaTFSI system. \newline
Additional interactions between ions and the polymer backbone are indicated by complementary $^{13}$C direct excitation solid-state NMR measurements with a short interscan delay of $1\,\mathrm{s}$, which are capturing predominantly mobile carbon moieties. A peak shifting from $71\,\mathrm{ppm}$ to lower ppm values upon addition of LiTFSI ($70.9\,\mathrm{ppm}$) or NaTFSI ($70.7\,\mathrm{ppm}$) together with a signal broadening of the PEG signal was observed. The corresponding NMR spectra are shown in \textbf{Figure \ref{fig:nmr}c}. Both effects are less pronounced in $^{13}$C cross-polarization spectra measured with a contact time of $2\,\mathrm{ms}$ highlighting more rigid carbon moieties. Consequently, the data inidcates that coordination of the cations seems to be occurring primarily with more mobile parts of the PEG chains, whereas rigid PEG domains are less influenced by addition of NaTFSI oder LiTFSI salt. For the two ions, the $^7$Li and $^{23}$Na NMR spectra show one specific kind of environment each (\textbf{Supp. Figure 11}). \newline
To gain additional insights into the underlying transport processes and their related timescales, impedance measurements were conducted on the solid polymer electrolytes as a function of NaTFSI salt concentration and temperature. The resulting Nyquist plots exhibit a semi-circle at all temperatures, along with a straight, inclined line towards lower frequencies (see \textbf{Supp. Figure 2}). Given a configuration where the hardened polymer electrolyte is positioned between two ion-blocking copper electrodes, impedance spectra for all samples were fitted using a Debye circuit including an additional series resistance, as illustrated in Supp. Figure 2 in the SI. Considering the cell geometry, the bulk conductivities were determined accordingly. At room temperature, the conductivity for sample Na2 ($c_{\mathrm{NaTFSI}}\,=\,0.2\,\mathrm{mol\,kg^{-1}}$) amounts to $\sigma\,=\,2.0\,\cdot\,10^{-4}\,\mathrm{S\,m^{-1}}$, i.e. slightly lower than that of the corresponding LiTFSI sample of the same salt concentration ($\sigma\,=\,4.0\,\cdot\,10^{-4}\,\mathrm{S\,m^{-1}}$ $^{[12]}$). Similar to the LiTFSI doped solid-state electrolytes, the temperature-dependent conductivity in NaTFSI-based samples, depicted in \textbf{Figure \ref{fig:seebeck}b}, can be described using the Vogel-Fulcher-Tammann (VFT) formalism $^{[12]}$:
\begin{equation}
\sigma \,=\,\sigma _{0}\,\cdot\,e^{\frac{-E_A}{k_B \left(T-T_{V}\right)}}
\end{equation}
where $T_V$ is the characteristic “Vogel temperature”, interpreted as temperature where the segmental ion conductivity drops to zero, and $E_A$ is the effective activation energy of ionic movement, respectively. \newline
As for the lithium samples, a concentration-dependent increase of activation energy can be observed. Thereby, the activation energy $E_A$ is increased from $21\,\mathrm{meV}$ without salt to $130\,\mathrm{meV}$ for a concentration of $c_{\mathrm{NaTFSI}}\,=\,0.5\,\mathrm{mol\,kg^{-1}}$, while the Vogel temperature declines to $T_V\,=\,\left(128\,\pm\,92\right)\,\mathrm{K}$. Thus, activation energy and Vogel temperature show an even more pronounced anticorrelated change compared to the neat polymer than the corresponding LiTFSI samples $^{[12]}$. \newline
By its relation to the glass transition temperature $^{[18]}$, the decline of the Vogel temperature upon salt addition can be interpretated by the softening of the matrix as $\mathrm{Li^+}$ as well as $\mathrm{Na^+}$ ions interfere with the crosslinking of the surrounding polymer-chains, thus leading to a “softening” of the polymer host. Again, this is corroborated by complementary solid-state NMR data. \newline
The two peaks, occurring in the $^1$H-NMR spectra at chemical shifts of $\delta\,=\,6.1\,\mathrm{ppm}$ and $\delta\,=\,6.6\,\mathrm{ppm}$ (see \textbf{Supp. Figure 4}) can be assigned to hydrogen atoms on unreacted methacrylate double bonds within the polymer backbone, thus implying that their relative integral is an estimate for the degree of crosslinking and, hence, the stiffness of the polymer. Although a precise quantification remains challenging, it is discernible that at equivalent LiTFSI and NaTFSI salt concentrations ($c\,=\,0.2\,\mathrm{mol\,kg^{-1}}$) and identical preparation protocols, the peaks are higher in amplitude than in the case of the neat polymer, with that of NaTFSI showing the highest value. A comparison of the $^{13}$C solid-state NMR data obtained by cross-polarization and direct excitation with short interscan delay also shows the presence of both more rigid and more mobile polymer fragments (see \textbf{Supp. Figures 11} and \textbf{12}). This observation coincides with the indication of a temperature dependent Vogel temperature and, therefore, supports that ion coordination prevents efficient formation of the polymer scaffold during UV-curing. \newline
The rise in activation energy $E_A$ can be attributed to an increased ion-ion $^{[17]}$ or ion-polymer interaction. Basically, this has a negative impact on the diffusion constant of the respective ionic charge carriers, resulting in a diminished mean conductivity $\sigma$. Nevertheless, an increase in conductivity with rising salt concentration can be observed, which is why it is conceivable that, overall, the additional morphological or configurational influences by the polymer matrix favor the mobility of ionic charge carriers. For instance, the influence of ions on the coordination of monomers within the polymer matrix seems to be plausible. Leading to an enhanced disorder in the polymer host, this mechanism could result in the formation of conductive percolation pathways, which positively affect the charge carrier diffusion within the electrolyte. \newline
Comparing the ionic conductivities $\sigma$ as function of the employed salt concentration (NaTFSI vs. LiTFSI) indicates that the polymer electrolyte hosting LiTFSI as conductive salt exhibits higher $\sigma$ values at identical concentrations and temperatures (see \textbf{Supp. Figure 5}). Obviously, the preferred embedding of $\mathrm{Li^+}$ ions in the polymer matrix results in a saturation of the polar groups at the polymer backbone even at lower concentrations compared to NaTFSI and thus to an increased contribution of freely moving ions which raise the conductivity accordingly.

\subsection{Impact of Carbon-Based Additives}
In $^{[12]}$ we demonstrated the tuning of the thermoelectric transport properties by adding multi-walled carbon nanotubes to the polymer blend. In the course of this investigation, we were able to identify a new mode of operation in which a thermoelectric generator (TEG) switches on autonomously above a certain ambient temperature due to the difference in electronic versus ionic conductivity as function of temperature, associated with a change in polarity of the thermovoltage.
To investigate the role of the additive’s dimensionality on the thermoelectric performance, we extended these studies on various, technologically relevant carbon-based additives. We selected single-walled carbon nanotubes (quasi-1D structures), graphite flakes (2D structures) and Carbon Black (3D structure) to investigate their impact on the thermoelectric transport properties, in particular the conductivity and the Seebeck coefficient. For this purpose, we used polymer blends with salt concentrations of $c\,=\,0.2\,\mathrm{mol\,kg^{-1}}$ LiTFSI or NaTFSI as references and mixed them with the respective additive.

\subsubsection*{Carbon Nanotubes (q1D)}
\begin{figure}[!h]
  \includegraphics[width=\linewidth]{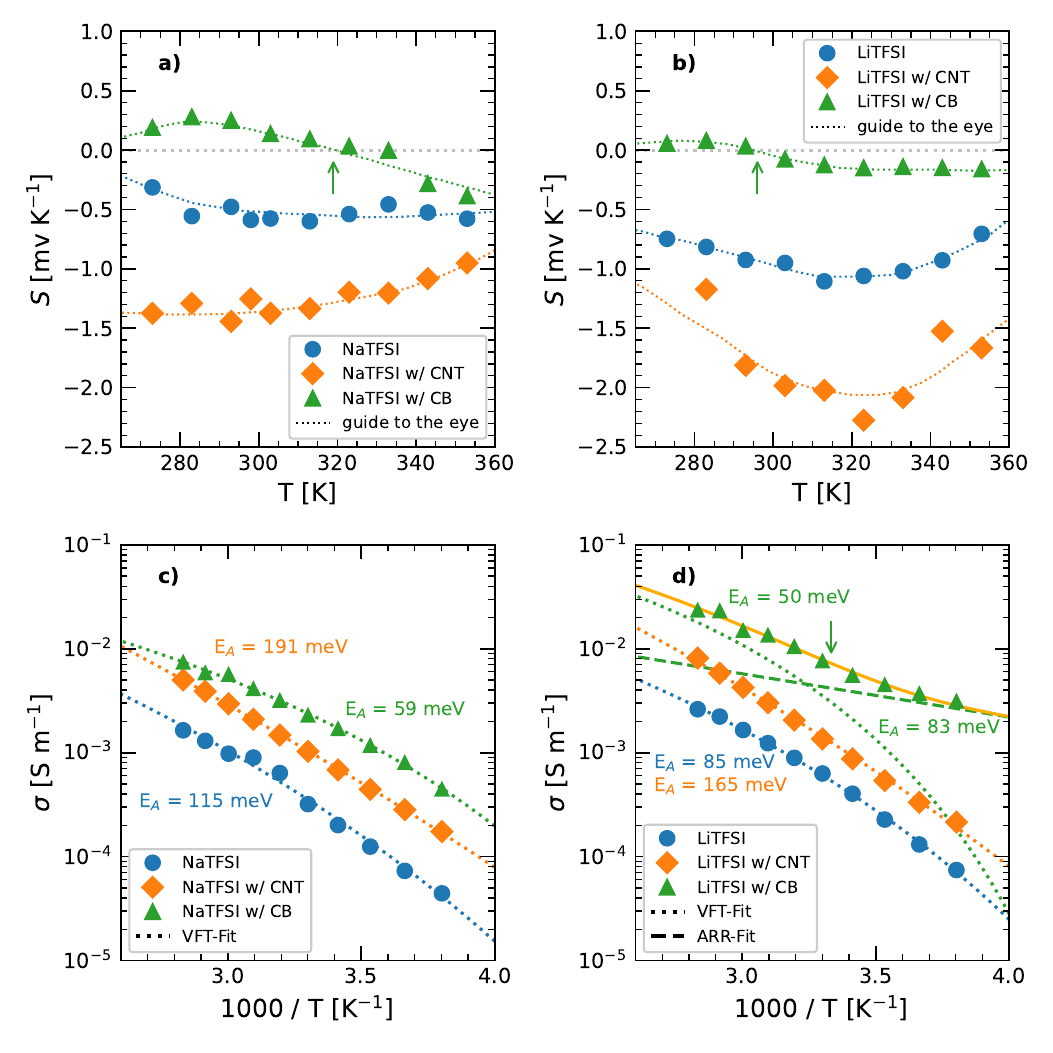}
  \caption{Seebeck coefficients of the neat electrolytes (blue dots) and the same electrolytes with carbon nanotubes (orange diamonds) or Carbon Black (green triangles) added to the polymer matrix containing $c\,=\,0.2\,\mathrm{mol\,kg^{-1}}$ of NaTFSI \textbf{(a)} and LiTFSI \textbf{(b)}, respectively. Samples containing Carbon Black show a change in polarity of the Seebeck coefficients at around $T_c\,=\,320\,\mathrm{K}$ (NaTFSI w/ CB) and $T_c\,=\,300\,\mathrm{K}$ (LiTFSI w/ CB), indicated by the arrows. A similar behaviour could be observed in samples containing multiwalled carbon nanotubes $^{[12]}$. Arrhenius plots of the conductivity for samples containing $c\,=\,0.2\,\mathrm{mol\,kg^{-1}}$ NaTFSI \textbf{(c)} and LiTFSI \textbf{(d)} salt (blue dots) as well as various additives: carbon nanotubes (orange diamonds) and Carbon Black (green triangles). The temperature dependent total conductivity $\sigma$ of the sample containing LiTFSI and Carbon Black can be modeled by overlapping Vogel-Fulcher-Tammann (dotted green line) and Arrhenius type (dashed green line) transport models ($\sigma\,=\,\sigma_{VFT}\,+\,\sigma_{Arr}$, yellow line), with VFT becoming dominant for higher temperatures $^{[12]}$. A change in the dominant transport mechanism is evident at around $T_c\,=\,300\,\mathrm{K}$.}
  \label{fig:transport}
\end{figure}

\begin{figure}[!h]
	\includegraphics[width=\linewidth]{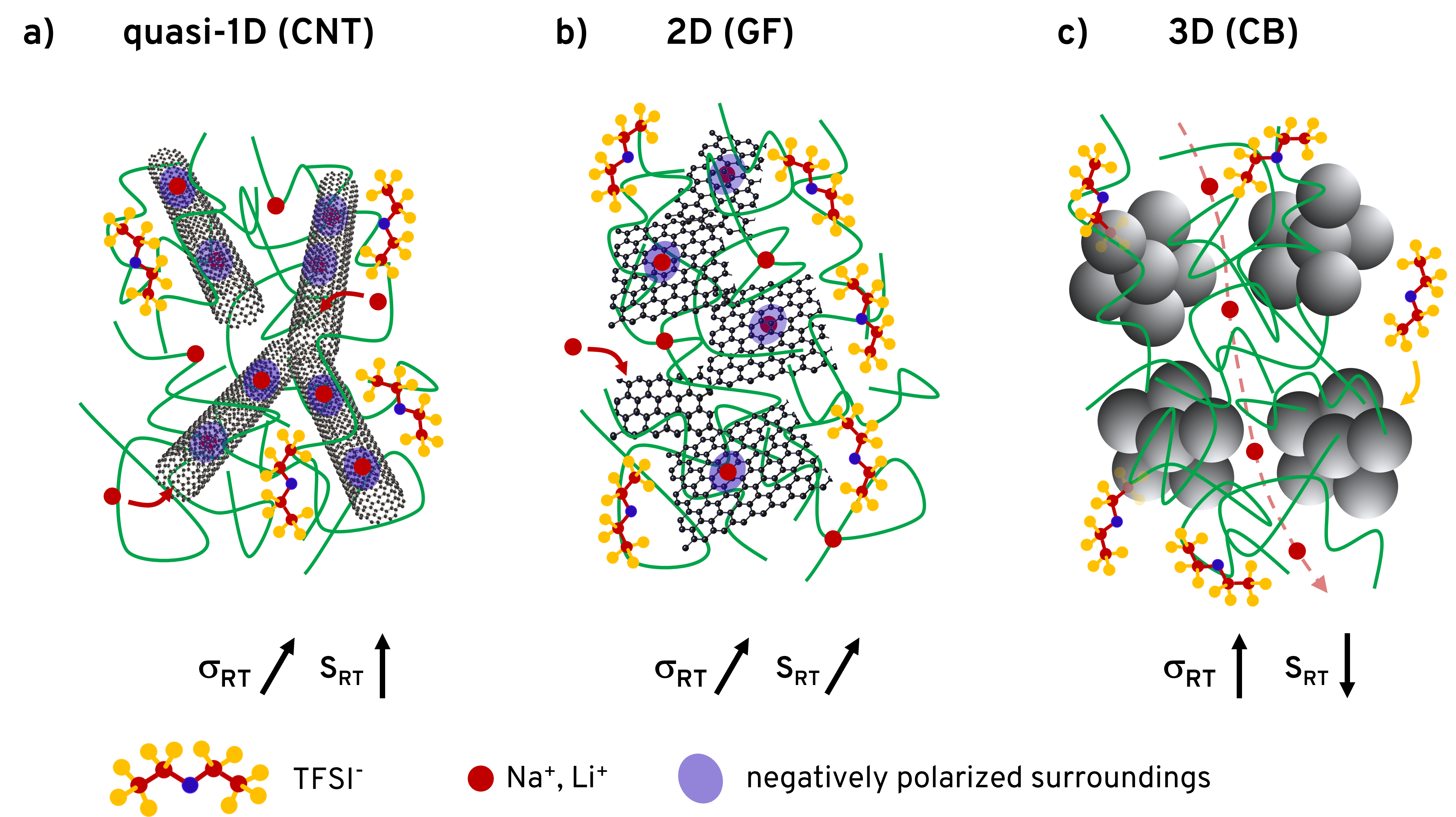}
	\caption{Effects of the different carbon-based additives and their respective dimension on the emerging transport and immobilization processes of cations and anions. \textbf{a}) In case of quasi-1D carbon nanotubes (CNTs) the formation of a percolation network within the polymer matrix results in an effective increase of electrical conductivity due to additional electronic transport contributions along the tubes and an increase of $\mathrm{TFSI^-}$ movement, presumably, as a result of directional alignment of polymers in proximity of the tubes. Simultaneously, a strong increase in the Seebeck coefficient results by the immobilization and infiltration of $\mathrm{Na^+}$ or $\mathrm{Li^+}$ cations by the CNTs. As an additional effect of the latter (not shown in the figure), the local polarization of the CNT surfaces results in an electrostatic interaction and thus, capacitive contribution that is confirmed by impedance spectroscopy data. \textbf{b}) For 2D graphene flakes, the formation of sheet percolation networks leads to an increase of $\sigma$ as well as, due to the immobilization of the cations, to an increase of $S$. As for CNT, the coordination of cations along the polymer strands leads to a change in crosslinking and hence stiffness of the host, which effectively influences also the motion of $\mathrm{TFSI^-}$ anions. Finally, \textbf{c}) carbon black particle additives lead to pronounced channeling of cations (red dashed arrow line) as well as to interaction with $\mathrm{TFSI^-}$ anions on their surface. Overall, the various microscopic mechanisms yield to the strongest increase of electrical conductivity $\sigma$ for all carbon additives, however, on the expenses of a strongly reduced Seebeck coefficient $S$ due to the more balanced transport of positive and negative ions in these composites.}
  \label{fig:comicTransport}
\end{figure}

Upon addition of carbon nanotubes (CNTs) to the polymer blends, a significant increase in the Seebeck coefficient can be observed in samples of both conductive salts (\textbf{Figure \ref{fig:transport} a} and \textbf{b}). At room temperature, the Seebeck coefficient increases by almost a factor of three from $S\,=\,-0.48\,\mathrm{mV\,K^{-1}}$ to $S\,=\,-1.39\,\mathrm{mV\,K^{-1}}$. Furthermore, it becomes evident that carbon nanotubes influence the temperature dependence of the Seebeck coefficient, rendering it overall more pronounced. This suggests that CNTs exert an influence on the mobility of ionic charge carriers, either by a reduction of $\mathrm{Na^+}$ mobility or by a direct improvement of $\mathrm{TFSI^-}$ mobility. Local surface charge potentials may induce an enhanced coordination or binding affinity towards sodium ions ($\mathrm{Na^+}$) (see \textbf{Figure \ref{fig:comicTransport} a}). An alternative rationale for the diminished mobility of $\mathrm{Na^+}$ ions is rooted in the unique cylindrical and porous architecture inherent to CNTs. Owing to their small ionic radius, $\mathrm{Na^+}$ ions are capable of infiltrating and accumulating within the intricate CNT structure.
To determine the conductivities from the impedance spectra, a modified equivalent circuit (see \textbf{Supp. Figure 6}) had to be used in order to account for the interactions between carbon nanotubes (CNTs) and $\mathrm{Na^+}$ ions as well as for the ion enrichment. In both cases, that is using LiTFSI or NaTFSI as conductive salt, the addition of the quasi-1D CNTs to the polymer blend led to an increase of the overall conductivity (compare \textbf{Figure \ref{fig:transport} c} and \textbf{d}). This can be attributed to the high electrical conductivity of the CNTs themselves facilitating charge carrier transport across the polymer matrix by the formation of conductive percolation pathways. It is also plausible that CNTs directly influence the alignment and coordination of monomers and consequently, polymer strands within the matrix, resulting in an enhanced conductivity.

\subsubsection*{Graphite Flakes (2D)}
Graphite flakes (GF) additives lead to a smaller increase in the absolute value of the Seebeck coefficient, see \textbf{Supp. Figure 8}. Compared to the additive-free samples Li2 or Na2, respectively, an increase in the magnitude of this quantity of about $0.5\,\mathrm{mV\,K^{-1}}$ can be observed. Again, this increase can be attributed to either a decrease in the $\mathrm{Na^+}$ ion mobility or an increase in the mobility of $\mathrm{TFSI^-}$ ions. GF dispersed into the polymer structure reduce the degree of polymer crosslinking, which facilitates ion diffusion through the polymer matrix (see \textbf{Figure \ref{fig:comicTransport} b}). 
Under the assumption that positively charged $\mathrm{Na^+}$ ions predominantly coordinate to the polar groups at the polymer backbones in the matrix, this increased mobility can still be attributed to a higher transport number for $\mathrm{TFSI^-}$ anions as confirmed by the increased, negative Seebeck coefficient. Another factor contributing to the higher Seebeck coefficient is the enhanced electronic contribution to the transport and thus to the thermoelectric voltage upon addition of GF.
Temperature dependent Nyquist plots for the samples containing sodium as conductive salt are shown in \textbf{Supp. Figure 7}. Under the limitation of an appropriate number of physical parameters to fit the curves, no straightforward equivalent circuit could be found to consistently describe the spectrum over the entire frequency range for all temperatures. To extract, at least, the sample resistance, the data was fitted with an R-CPE parallel circuit, which approximates the minima of the impedance curves adequately. \textbf{Supp. Figure 9} visualizes that the electrical conductivity $\sigma$ increases with the addition of GF compared to the neat polymer samples Li2 and Na2. Hence, it is likely that GF additives form an embedded, conductive network within the polymer host that sustains charge transport across the electrolyte. Here, the 2D laterally extended graphite structure of the GF could act as conductive bridge across isolated material defects or as link between short-range ordered polymer domains, significantly facilitating charge carrier transport by thermally activated hopping processes.

\subsubsection*{Carbon Black (3D)}
When Carbon Black (CB), characterized by its 3D para-crystalline structure in combination with a high surface-to-volume ratio, is used as additive in the solid electrolyte, its Seebeck coefficient exhibits a distinctive behavior within the measured temperature range. Up to a temperature of $320\,\mathrm{K}$, the Seebeck coefficient possesses a positive sign. Subsequently, it crosses the zero point and changes its sign to negative, with its magnitude continuously increasing with rising temperature (Figure \ref{fig:transport} a and b). These observations suggest that the mobility of $\mathrm{Na^+}$ ions, unlike in two-dimensional GF or quasi-one-dimensional CNTs, is much easier thermally activated in the range from $273\,\mathrm{K}$ to $310\,\mathrm{K}$ than that of the $\mathrm{TFSI^-}$ counterions. This could originate from either the stronger immobilization of $\mathrm{TFSI^-}$ anions in combination with a higher energy barrier for relaxation or by the significant mobilization of $\mathrm{Na^+}$ ions in presence of the CB additive. It is conceivable that $\mathrm{TFSI^-}$ ions penetrate the layered CB structure and accumulate in its volume, like $\mathrm{Na^+}$ ions in CNTs. This would lead to a higher concentration of freely moving $\mathrm{Na^+}$ ions which ultimately would determine the thermoelectric voltage in the specified temperature range. Above a critical temperature, $\mathrm{TFSI^-}$ ions may be thermally released from the para-crystalline CB planes and leave the additive’s volume, resulting in an excess of mobile $\mathrm{TFSI^-}$ ions compared to $\mathrm{Na^+}$ ions and, in this scenario, constituting the primary contribution to the thermoelectric voltage. This would account for the observed change in sign change of the Seebeck coefficient. \newline
Impedance spectra of samples Li2CB and Na2CB are fitted based on  the electrical equivalent circuit shown in \textbf{Supp. Figure 10}. Parameters CPE2 and R2 account for the capacitive behavior originating by the $\mathrm{TFSI^-}$ enrichment and the resistance caused by the formed percolation pathways as well as the inherent resistance of the CB particles. Due to the small interlayer spacing of less than a $1\,\mathrm{nm}$ $^{[19][20]}$ between the para-crystalline graphite layers in the CB particles, the polymer electrolyte cannot penetrate into the primary particles and their aggregates. \newline
An increase in electrolyte conductivity can be observed for both, LiTFSI as well as NaTFSI conductive salt, see Figure 3c and d. As for GF and CNTs additives, the formation of conductive networks by the CB particles, their aggregates and agglomerations appears plausible. Overall, this would facilitate charge carrier motion by channeling of cations and hence increases the conductivity. \textbf{Figure 4c} visualizes the underlying concept. The internal structure of the primary Carbon Black particles offers a vast enhancement of the surface area being responsible for ion-CB as well as polymer-CB interaction. This is caused by the high porosity and the large number of defect sites within the particles. However, for the LiTFSI doped polymer electrolyte a kink occurs in the temperature dependent conductivity data. This has been observed also in polymer electrolytes based on LiTFSI and multiwalled CNT additives before $^{[12]}$ and was attributed to two superimposed transport processes in the respective temperature ranges, namely ionic and polymeric segmental mobility. \newline
The influence of the investigated additives on both the conductivity ($\sigma$) and the Seebeck coefficient ($S$) of the polymer electrolyte hints at a common underlying mechanism, although notable distinctions in the related datasets of these two parameters can be identified. Carbon nanotubes (CNTs) added to the polymer matrix result in the highest absolute Seebeck coefficients. Their structural characteristics support an enhanced accumulation of sodium ions ($\mathrm{Na^+}$) when compared to bulky Carbon Black (CB) or planar graphite flakes (GF), consequently diminishing the mobility of the $\mathrm{Na^+}$ species. Nevertheless, it is noteworthy that CNTs display the least marked increase in conductivity compared to CB and GF, presumably attributed to the concentration of CNTs within the polymer electrolyte. While the CNT concentration ($0.004\,wt\%$) approximates that of CB ($0.007\,wt\%$), CB particles offer a substantially larger effective surface area available for ion interaction and intercalation than CNT additives. Consequently, one can infer that CB exerts a more substantial influence on the increase in the overall charge carrier density. The reduced Vogel temperatures in all samples hosting carbon-based additives suggest that the latter influence the structural relaxation mechanisms of the polymer electrolyte. Furthermore, the reduction in the efficiency of the polymerization reaction leading to less connected polymer monomers contributes to the development of $T_V$. \newline
When comparing graphite flakes (GF) to Carbon Black (CB), CB demonstrate superior conductivity owing to its notably larger surface area. This increase in surface area results in a higher charge carrier density. Finally, one has to take into account that the 3D morphology of Carbon Black particles is much more in favor of forming conductive percolation paths across the solid electrolyte compared to the planar graphite flakes, which are preferentially aligned in a spatially isotropic orientation and do not show such a tendency to agglomerate. Hence, CB tends to form bigger joined conductive networks compared to GF.

\section{Conclusion}
The thermoelectric properties of solid polymer electrolytes were investigated as function of lithium bis(trifluoromethanesulfonyl)imid (LiTFSI) and sodium bis(trifluoromethanesulfonyl)imide (NaTFSI) conductive salt concentration as well as in dependence of carbon-based additives of different dimensionality (1D carbon nanotubes, 2D graphite flakes, and 3D Carbon Black). The results revealed that with increasing salt concentration, the Seebeck coefficients increase up to a distinct temperature, primarily caused by the incremental number of free charge carriers per volume and the saturation of this effect by either relaxation of all accessible charges or repulsive charge effects impeding the further increase in conductivity.
In the salt-only-electrolytes, a negative Seebeck coefficient is measured for all concentrations, implying that the TFSI- ions are the dominant mobile ionic species. The Vogel-Fulcher-Tammann model reveals an increasing activation energy $E_A$ with decreasing Vogel temperature $T_V$, the latter related to the polymer glass transition temperature and hence indicating changes in the polymerization upon increasing salt concentration, which is supported by complementary solid-state NMR measurements. Carbon nanotubes significantly enhance the Seebeck coefficient and its temperature dependence by influencing ion mobility and coordination. Graphite flakes only moderately enhance the Seebeck coefficient by decreasing the $\mathrm{Na^+}$ ion mobility and increasing the $\mathrm{TFSI^-}$ ion mobility, as well as contributing to electronic conductivity. Carbon Black exhibits a unique behavior, where the sign of the Seebeck coefficient changes with temperature, presumably due to an influence on the $\mathrm{Li^+}$ and $\mathrm{Na^+}$ ion mobility by interactions with $\mathrm{TFSI^-}$ ions residing within the additive’s structure. Temperature dependent conductivity measurements show two superimposed transport mechanisms, occurring upon adding CB in the lithium-based solid polymer electrolyte. The study also highlighted changes in activation energy and Vogel temperature with salt concentration, indicating structural modifications in the polymer matrix as function of the respective additive. Overall, the choice of carbon-based additives, and in particular their dimensionality and internal structure, as well as the salt concentration significantly influences the thermoelectric properties of the polymer electrolytes, providing insights into their potential and further optimization for thermoelectric applications. Our studies indicate that sodium, as a substitute for lithium in electrolyte salts, only minimally affects the thermoelectric properties, but due to its other positive aspects, represents a potentially more sustainable material system than lithium, thus showing great technological future potential for electrolyte-based thermoelectric applications, as also confirmed by current trends and developments in battery research.

\section{Experimental Section}
\threesubsection{Materials and sample preparation}\\
Poly (ethylene glycol) methyl ether methacrylate (PEGMA, Mn = 500) and bisphenol A ethoxylate di-methacrylate (BEMA, Mn = 1700) were obtained from Sigma Aldrich. Prior to use, they were kept in the inert atmosphere of an Ar-filled dry glove box for a couple of days to allow for sublimation of volatile contaminants. PEGMA and BEMA were mixed in a 3:7 weight ratio. Different contents of LiTFSI (Sigma Aldrich) and NaTFSI (TCI Chemicals), respectively, were stirred into the polymer blend. Various carbon-based additives (Carbon Black, Graphite Flakes and Carbon Nanotubes (SG65i) (Sigma Aldrich)) were chosen for the thermoelectric investigations. (6,5)-SWCNT enriched CNTs were mixed into PEGMA and sonicated using a SONIFIER 450 (Branson) by Y. Vollert, Physical Chemistry II, Julius-Maximilians-Universität Würzburg, after purchase. PEGMA (incl. CNTs) and BEMA were later processed as described above. $2–4\,\mathrm{wt\%}$ of 2-hydroxy-2-methylpropiophenone was added as photo initiator. The polymer was filled in a test cell and solidified by UV curing ($\lambda\,=\,400\,\mathrm{nm}$) under inert gas atmosphere.

\threesubsection{Determination of transport properties}\\
Ionic and electronic conductivities of the solid polymer electrolyte (SPE) were measured by means of impedance spectroscopy in vertical cells of copper  $||$ SPEs $||$ copper using a Zurich Instruments MFIA impedance analyzer in a frequency range from $100\,\mathrm{mHz}$ to $510\,\mathrm{kHz}$. Experimental data were gathered in a temperature range between $263$ and $363\,\mathrm{K}$. Impedance spectra were fitted using electrical equivalent circuits and conductivities were calculated according to its definition $\sigma\,=\,L\,/\,\left(A\,\cdot\,R\right)$ with $L$ and $A$ being thickness and area of the samples. $R$ denotes the resistance, which was obtained from the respective impedance spectra. Thermoelectric measurements were performed on vertical cells composed of copper $||$ SPE $||$ copper in which the two contact sides could be heated/cooled independently using Peltier devices (Telemeter Electronics). By varying the temperature around a constant mean value, the resulting gradient give rise to the thermovoltage, which was recorded with a Keithley KE2182 A nanovoltmeter. From the slope of the measured thermovoltage ($U_{th}$) versus the applied temperature difference ($\Delta T$) curve, the Seebeck coefficient can be determined $^{[21]}$. Measurements are carried-out continuously under quasi-steady state conditions. Heating rates of $1\,\mathrm{K\,min^{-1}}$ and holding times of $30\,\mathrm{s}$ at a given temperature difference assure sufficient time to access the steady-state thermovoltage $^{[22]}$. The employed temperature difference was limited to $\Delta T_{max}\,=\,10\,\mathrm{K}$. The data obtained were corrected for the Seebeck coefficient of the contacting wires (made of Cu) which is well documented in literature $^{[23]}$ ($S_{Cu}\,=\,1.5\,\mathrm{\mu V\,K^{-1}}$ at room temperature). All measurements mentioned above were carried out under vacuum conditions ($p\,\approx\,1–10\,\mathrm{mbar}$).

\threesubsection{Solid-state NMR}\\
Solid-state NMR experiments were performed on a Bruker Avance Neo NMR Spectrometer at $9.4\,\mathrm{T}$ with either $8\,\mathrm{kHz}$ or $9\,\mathrm{kHz}$ MAS. A $4\,\mathrm{mm}$ HX double channel probe was used. The $90^{\circ}$ pulse durations for $^1$H, $^7$Li, $^{19}$F and $^{23}$Na were $2.50\,\mathrm{\mu s}$, $4.0\,\mathrm{\mu s}$, $2.55\,\mathrm{\mu s}$ and $4.0\,\mathrm{\mu s}$, respectively. $^1$H-$^{13}$C cross-polarization (CP) experiments with contact times of $0.05\,\mathrm{ms}$, $0.5\,\mathrm{ms}$ and $2\,\mathrm{ms}$ with a ramp on the $^1$H channel were performed for each sample. During acquisition, SPINAL64 heteronuclear decoupling with $100\,\mathrm{kHz}$ was employed. $^{13}$C NMR spectra with direct excitation (DE) were recorded with a short interscan delay of $1\,\mathrm{s}$. Chemical shifts were referenced using adamantane. All experiments were nominally performed at RT. However, frictional heating due to MAS results in a higher actual sample temperature. Spectral processing was done using the Bruker TopSpin 4.1.4 software.

\medskip
\textbf{Supporting Information} \par 
Supporting Information is available from the Wiley Online Library or from the author.

\medskip
\textbf{Acknowledgements} \par
MF, JSS \& JP thank Y. Vollert and T. Hertel, Physical Chemistry II, Julius-Maximilians-Universität Würzburg, for providing PEGMA soluted CNTs and the Bavarian Ministry of Science and the Arts for the generous support by the research program \textit{Solar Technologies Go Hybrid}.

\medskip
\textbf{Conflict of Interest} \par
The authors declare no conflict of interest.

\medskip
\textbf{Data availability statement} \par 
The data that support the findings of this study are available from the corresponding author upon reasonable request.

\medskip
\textbf{References} \par \noindent
{[1]}	L. Long, S. Wang, M. Xiao, Y. Meng, \textit{J. Mater. Chem. A} \textbf{2016}, \textit{4}, 10038.\\
{[2]}	C. Forman, I. K. Muritala, R. Pardemann, B. Meyer, \textit{Renewable Sustainable Energy Rev.} \textbf{2016}, \textit{57}, 1568.\\
{[3]}	G. J. Snyder, E. S. Toberer, \textit{Nat. Mater.} \textbf{2008}, \textit{7}, 105.\\
{[4]}	M. Goel, M. Siegert, G. Krauss, J. Mohanraj, A. Hochgesang, D. C. Heinrich, M. Fried, J. Pflaum, M. Thelakkat, \textit{Adv. Mater.} \textbf{2020}, \textit{32}, 2003596.\\
{[5]}	T. Ma, W. Kent, B. X. Dong, G. L. Grocke, S. N. Patel, \textit{Appl. Phys. Lett.} \textbf{2021}, \textit{119}, 013902.\\
{[6]}	F. Huewe, A. Steeger, K. Kostova, L. Burroughs, I. Bauer, P. Strohriegl, V. Dimitrov, J. Pflaum, \textit{Adv. Mater.} \textbf{2017}, \textit{29}, 1605682.\\
{[7]}	I. Petsagkourakis, K. Tybrandt, X. Crispin, I. Ohkubo, N. Satoh, T. Mori, \textit{Sci. Technol. Adv. Mater.} \textbf{2018}, \textit{19}, 836.\\
{[8]}	O. Bubnova, Z. U. Khan, A. Malti, S. Braun, M. Fahlmann, M. Berggren, X. Crispin, \textit{Nat. Mater.} \textbf{2011}, \textit{10}, 429.\\
{[9]}	J. Wüsten, K. Potje-Kamloth, \textit{J. Phys. D: Appl. Phys.} \textbf{2008}, \textit{41}, 135113.\\
{[10]}	H. Wang, U. Ail, R. Gabrielsson, M. Berggren, X. Crispin, \textit{Adv. Energy Mater.} \textbf{2015}, \textit{5}, 1500044.\\
{[11]}	M. Bharti, A. Singh, A. K. Debnath, A. K. Chauhan, K. P. Muthe, S. K. Gupta, K. Marumoto, T. Mori, D. K. Aswal, \textit{Mater. Today Phys.} \textbf{2021}, \textit{16}, 100307.\\
{[12]}	M. Frank, J. Pflaum, \textit{Adv. Funct. Mater.} \textbf{2022}, \textit{32}, 2203277.\\
{[13]}	D. Zhao, A. Würger, X. Crispin, \textit{J. Energy Chem.} \textbf{2021}, \textit{61}, 88.\\
{[14]}	J. R. Nair, C. Gerbaldi, M. Destro, R. Bongiovanni, N. Penazzi, \textit{React. Funct. Polym.} \textbf{2011}, \textit{71}, 409.\\
{[15]}	M. D. Slater, D. Kim, E. Lee, C. S. Johnson, \textit{Adv. Funct. Mater.} \textbf{2013}, \textit{23}, 947.\\
{[16]}	X. Wang, Q. Zhang, C. Zhao, H. Li, B. Zhang, G. Zeng, Y. Tang, Z. Huang, I. Hwang, H. Zhang, S. Zhou, Y. Qiu, Y. Xiao, J. Cabana, C.-J. Sun, K. Amine, Y. Sun, Q. Wang, G.-L. Xu, L. Gu, Y. Qiao, S.-G. Sun, \textit{	Nat. Energy} \textbf{2024}.\\
{[17]}	M. Takeuchi, Y. Kameda, Y. Umebayashi, S. Ogawa, T. Sonoda, S.-I. Ishiruo, M. Fujita, M. Sano, \textit{J. Mol. Liq.} \textbf{2009}, \textit{148}, 99.\\
{[18]}	J. Mindemark, M. J. Lacey, T. Bowden, D. Brandell, \textit{Prog. Polym. Sci.} \textbf{2018}, 114.\\
{[19]}	L. Spanu, S. Sorella, G. Galli, \textit{Phys. Rev. Lett.} \textbf{2009}, \textit{103}, 196401.\\
{[20]}	M. Ue, \textit{J. Electrochem. Soc.} \textbf{1994}, \textit{141}, 3336.\\
{[21]}	J. de Boor, E. Müller, \textit{Rev. Sci. Instrum.} \textbf{2013}, \textit{84}, 065102.\\
{[22]}	K. A. Borup, J. de Boor, H. Wang, F. Drymiotis, F. Gascoin, X. Shi, L. Chen, M. I. Fedorov, E. Müller, B. B. Iversen, G. J Snyder, \textit{Energy Environ. Sci.} \textbf{2015}, \textit{8}, 423.\\
{[23]}	F. J. Blatt, R. H. Kropschot, \textit{Phys. Rev.} \textbf{1960}, \textit{118}, 480.\\
{[24]}	J. R. Nair, C. Gerbaldi, G. Meligrana, R. Bongiovanni, S. Bodoardo, N. Penazzi, P. Reale, V. Gentilli, \textit{J. Power Sources} \textbf{2008}, \textit{178}, 751.\\

\newpage

\cleardoublepage



\section*{Supplementary material}

# Supplementary Information

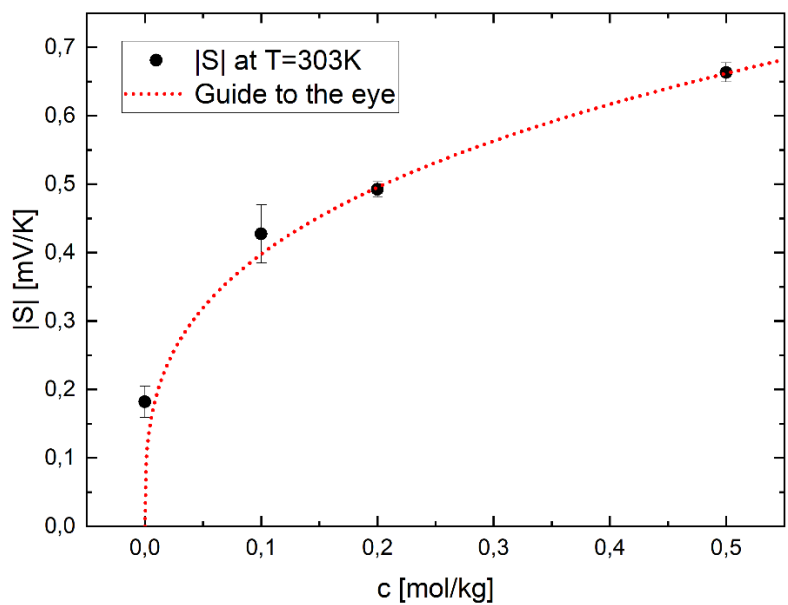

***Supp. Figure 1*** *Non-linear relationship of the absolute magnitude of the Seebeck coefficient with respect to the salt concentration at T=303 K.*

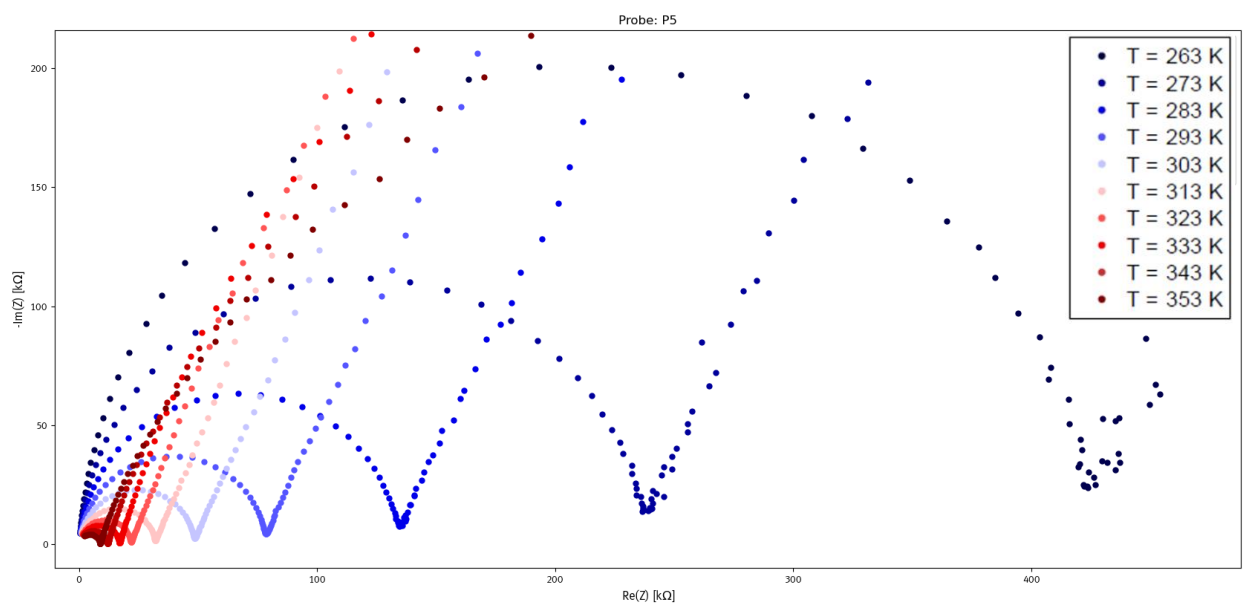

***Supp. Figure 2*** *Nyquist plots measured at a sample containing $c_{NaTFSI} = 0.5\ mol\ kg^{-1}$ salt. Impedance measurements were performed in a temperature range from 263 K to 353 K. The impedance spectra feature a characteristic semi-circle with a straight but inclined line towards lower frequencies.*

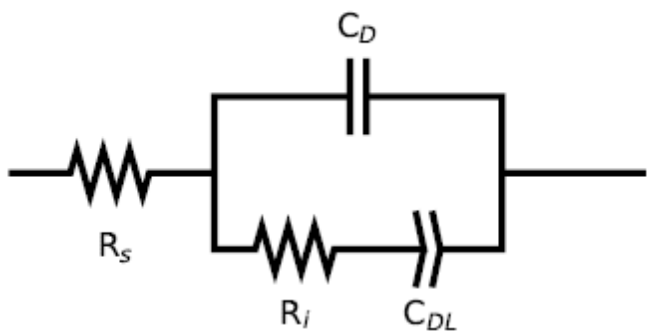

**Supp. Figure 3** *Equivalent electrical circuit used to fit the impedance measurements.*

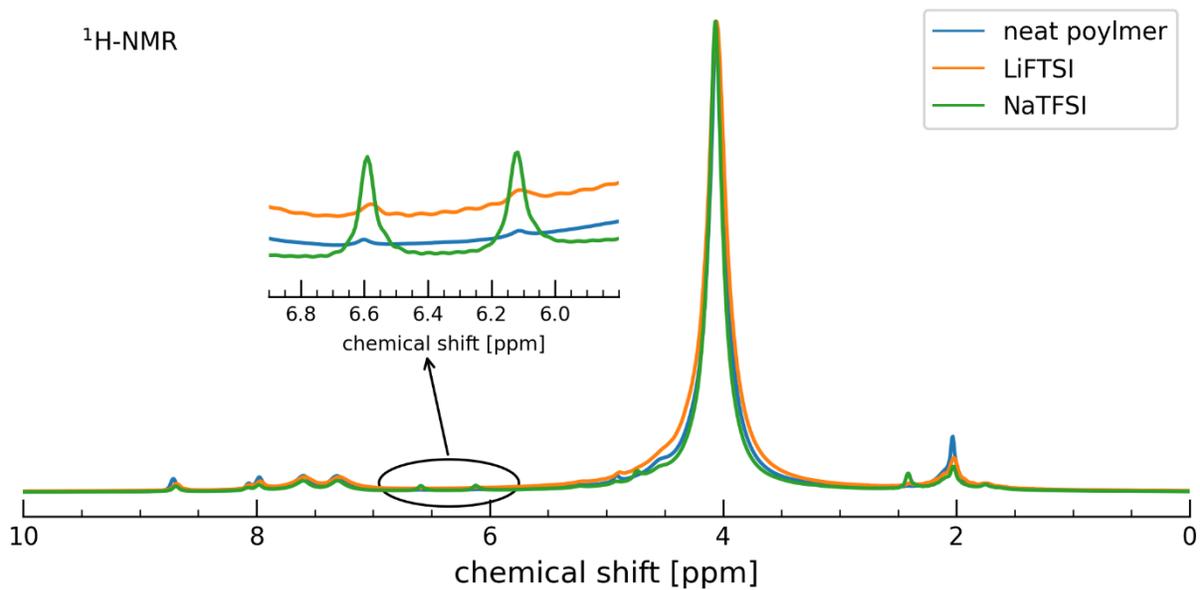

**Supp. Figure 4** *$^1$H-NMR measurement. Inset shows the signal of the carbonyl group located at the polymer backbone.*

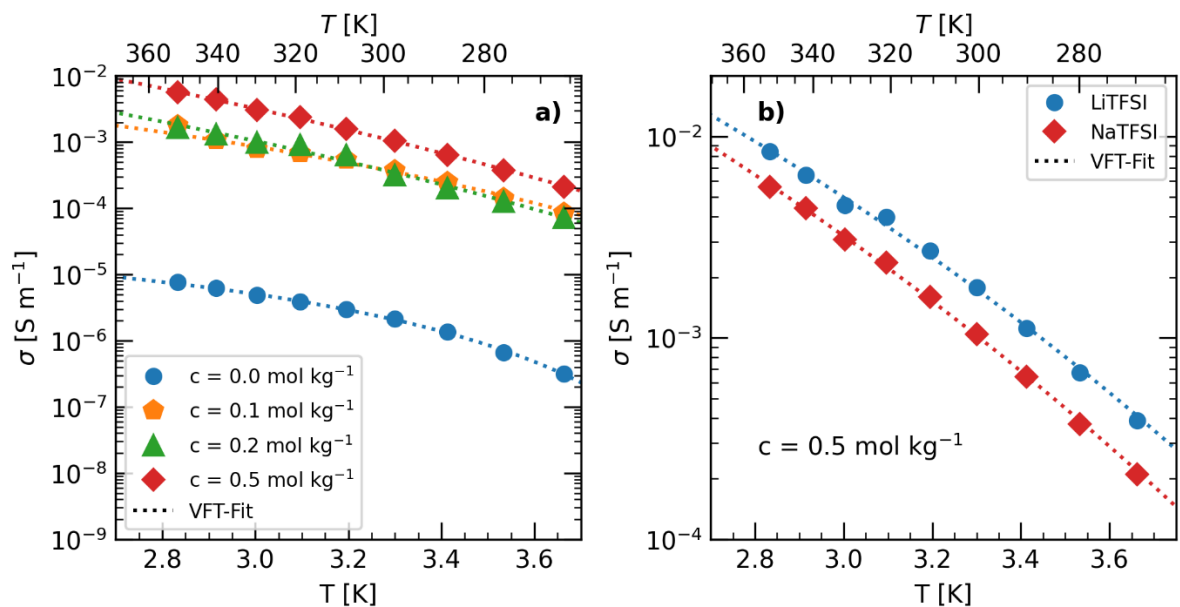

**Supp. Figure 5 a)** Temperature dependent conductivities for various concentrations of NaTFSI. **b)** Temperature dependent conductivities for samples containing $c = 0.5\ mol\ kg^{-1}$ LiTFSI and NaTFSI salt, respectively.

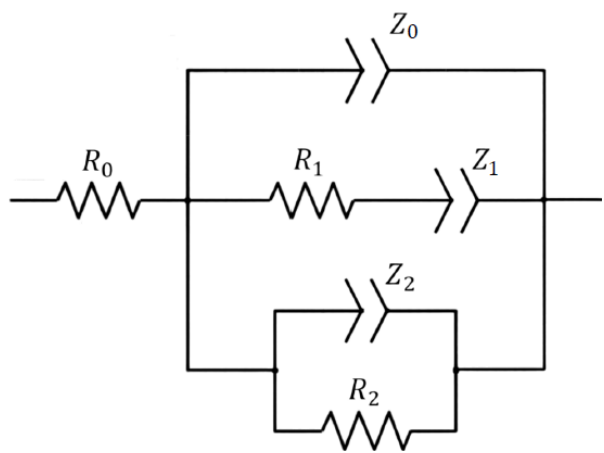

**Supp. Figure 6** Equivalent electrical circuit used to describe the impedance spectra of samples containing carbon nanotubes.

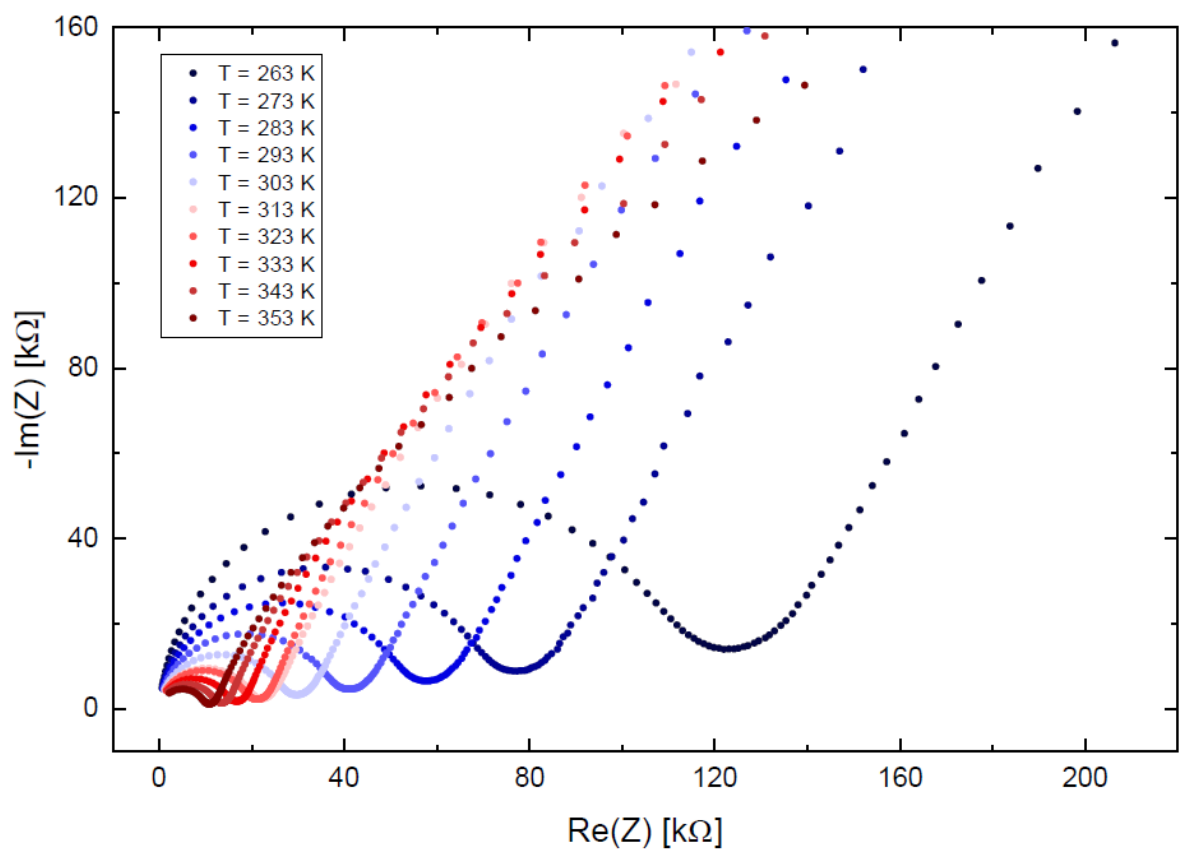

**Supp. Figure 7** Temperature dependent Nyquist plots of a polymer electrolyte containing graphite flakes.

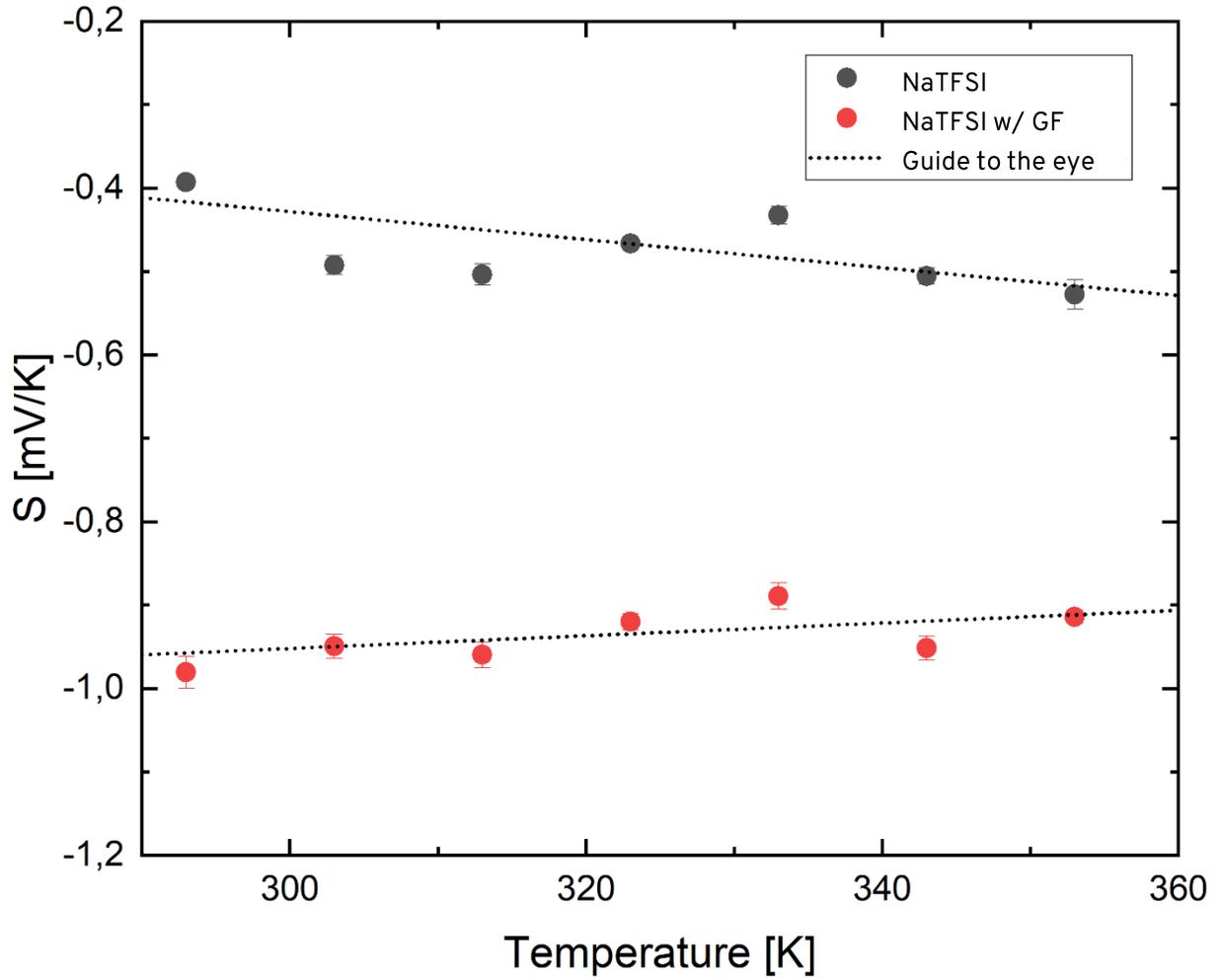

***Supp. Figure 8*** *Comparison of the Seebeck coefficients of samples containing NaTFSI conductive salt at a concentration of $c_{\text{NaTFSI}} = 0.2 \ mol \ kg^{-1}$ (black dots). The addition of graphite flakes (GF, red dots) increases $|S|$ by about $0.5 \ mV$, i.e. a factor of two.*

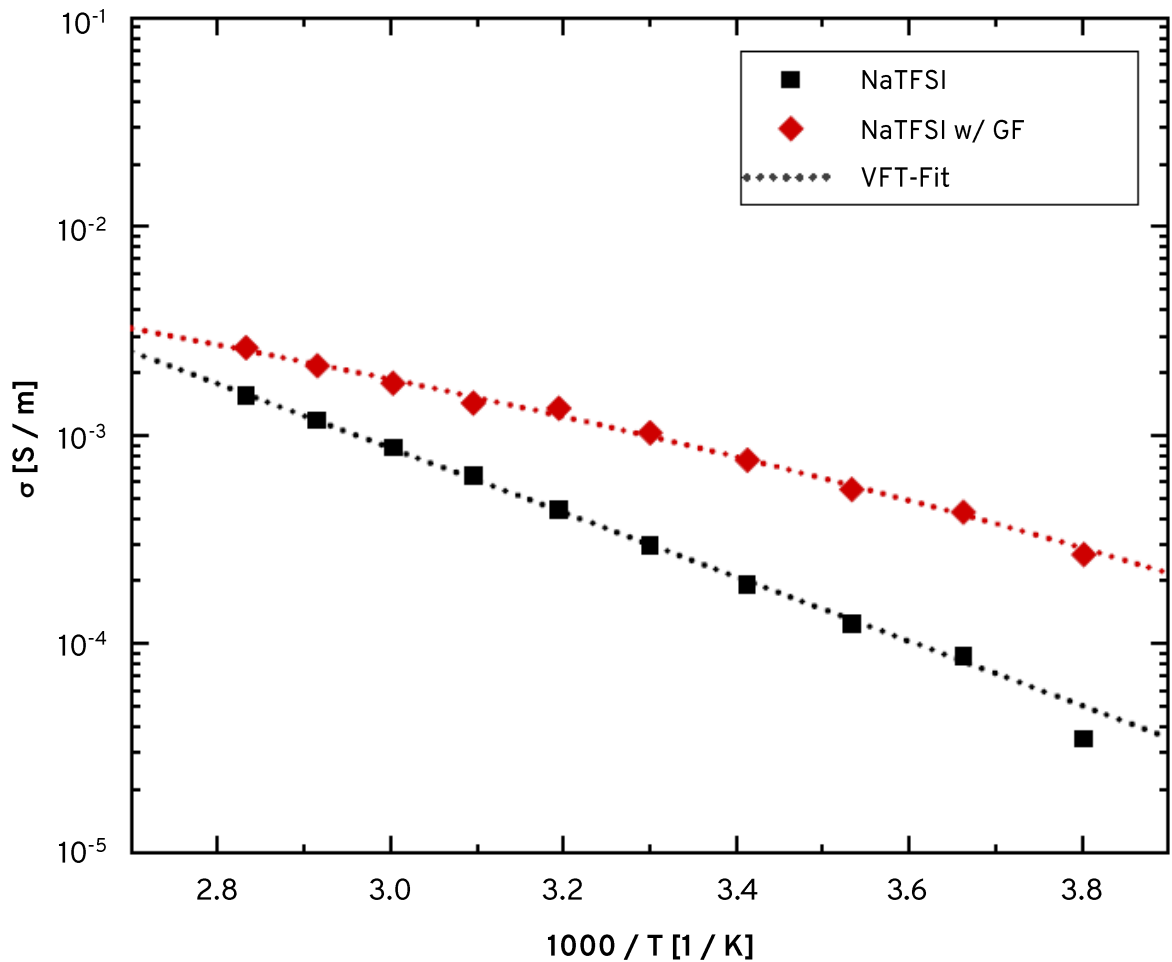

**Supp. Figure 9** Temperature dependent conductivities for samples containing NaTFSI conductive salt at a concentration of $c_{\text{NaTFSI}} = 0.2\ mol\ kg^{-1}$ as well as graphite flakes (red diamonds) compared to the respective data for the electrolyte with the same salt concentration but without GF additive (black squares). Dotted lines are fits according to the Vogel-Fulcher-Tammann (VFT) formalism.

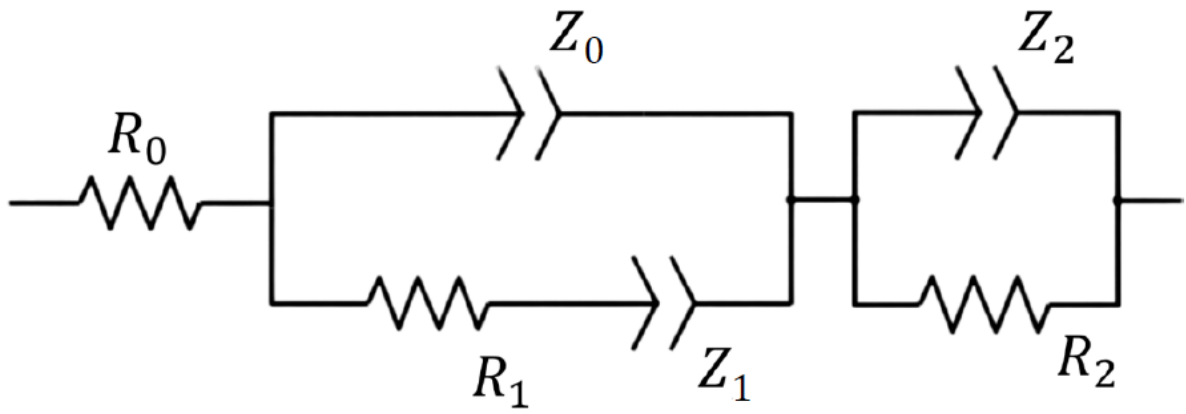

Supp. Figure 10 Equivalent electrical circuit used to describe the impedance spectra of samples containing carbon black.

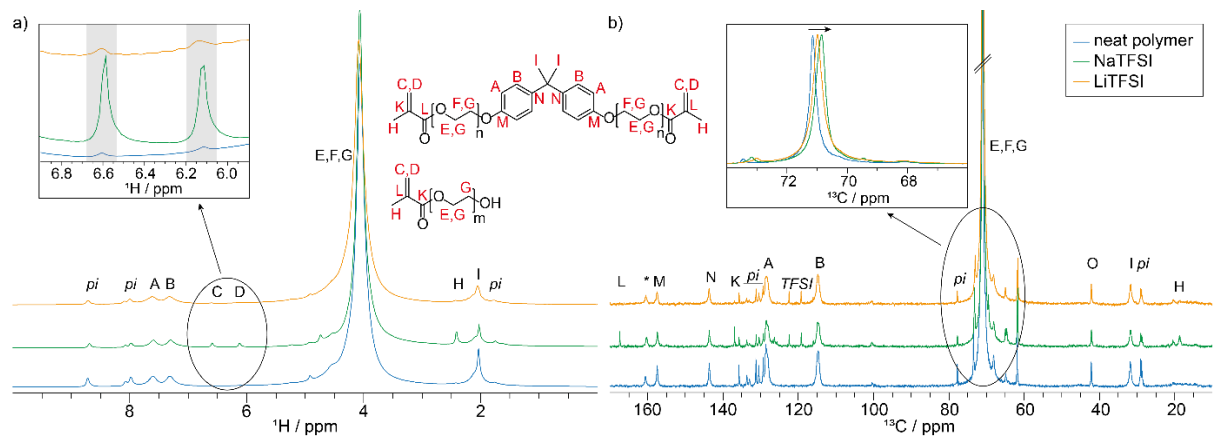

***Supp. Figure 11*** *Solid-state NMR spectra of the neat polymer (blue), the polymer with NaTFSI (green) and the polymer with LiTFSI (orange) together with the uncured polymer structure for peak assignment.* ***a)*** *Overlay of the $^1$H NMR spectra with peak assignment (pi = photoinitiator, * = rotational side band) and an additional zoom into the spectral region of the methacrylate double bond.* ***b)*** *Overlay of the $^{13}$C NMR spectra recorded with direct excitation and a short interscan delay of 1 s showing predominantly mobile carbons including peak assignment. An additional zoom into the PEG backbone carbon signals, which are scaled to matching peak heights, shows the peak changes upon addition of the different TFSI salts.*

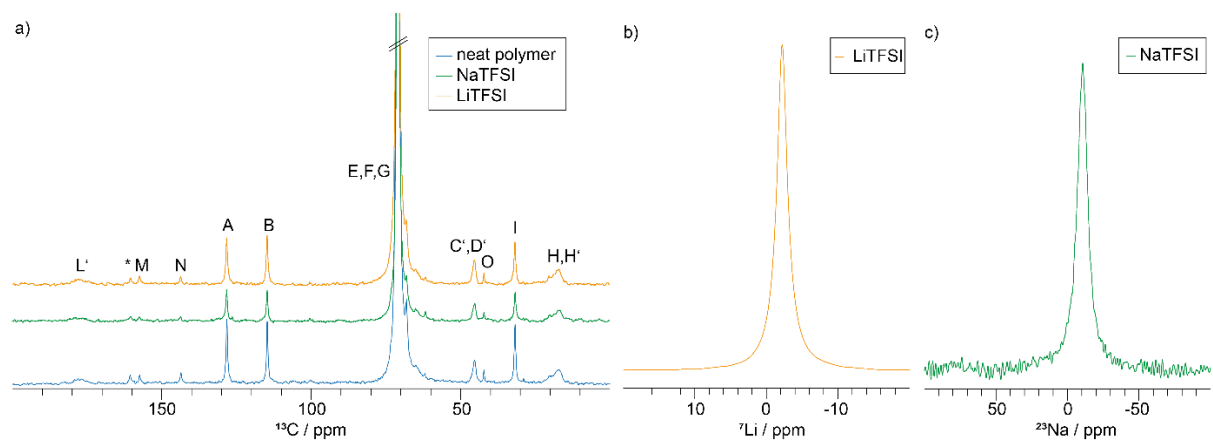

***Supp. Figure 12*** *Solid-state NMR spectra of the neat polymer (blue), the polymer with NaTFSI salt (green) and the polymer with LiTFSI salt (orange).* ***a)*** *Overlay of the $^{13}$C CP MAS NMR spectra with peak assignment.* ***b)*** *$^7$Li NMR spectrum of the polymer with the LiTFSI salt.* ***c)*** *$^{23}$Na NMR spectrum of the polymer with NaTFSI.*



\end{document}